\documentstyle[galley,epsf]{mn}
\newif\ifAMStwofonts

\ifoldfss
  \ifCUPmtlplainloaded \else
    \NewTextAlphabet{textbfit} {cmbxti10} {}
    \NewTextAlphabet{textbfss} {cmssbx10} {}
    \NewMathAlphabet{mathbfit} {cmbxti10} {} 
    \NewMathAlphabet{mathbfss} {cmssbx10} {} 
  \fi
  \ifAMStwofonts
    \ifCUPmtlplainloaded \else
      \NewSymbolFont{upmath} {eurm10}
      \NewSymbolFont{AMSa} {msam10}
      \NewMathSymbol{\upi}     {0}{upmath}{19}
      \NewMathSymbol{\umu}     {0}{upmath}{16}
      \NewMathSymbol{\upartial}{0}{upmath}{40}
      \NewMathSymbol{\leqslant}{3}{AMSa}{36}
      \NewMathSymbol{\geqslant}{3}{AMSa}{3E}

    \fi
  \fi
\fi 

\ifnfssone
  \newmathalphabet{\mathit}
  \addtoversion{normal}{\mathit}{cmr}{m}{it}
  \addtoversion{bold}{\mathit}{cmr}{bx}{it}
  \newmathalphabet{\mathbfit} 
  \addtoversion{normal}{\mathbfit}{cmr}{bx}{it}
  \addtoversion{bold}{\mathbfit}{cmr}{bx}{it}
  \newmathalphabet{\mathbfss} 
  \addtoversion{normal}{\mathbfss}{cmss}{bx}{n}
  \addtoversion{bold}{\mathbfss}{cmss}{bx}{n}
  \ifAMStwofonts
    \ifCUPmtlplainloaded \else
      \UseAMStwoboldmath
      \makeatletter
      \new@mathgroup\upmath@group
      \define@mathgroup\mv@normal\upmath@group{eur}{m}{n}
      \define@mathgroup\mv@bold\upmath@group{eur}{b}{n}
      \edef\UPM{\hexnumber\upmath@group}
      \new@mathgroup\amsa@group
      \define@mathgroup\mv@normal\amsa@group{msa}{m}{n}
      \define@mathgroup\mv@bold\amsa@group{msa}{m}{n}
      \edef\AMSa{\hexnumber\amsa@group}
      \makeatother
      \mathchardef\upi="0\UPM19
      \mathchardef\umu="0\UPM16
      \mathchardef\upartial="0\UPM40
      \mathchardef\leqslant="3\AMSa36
      \mathchardef\geqslant="3\AMSa3E
    \fi
  \fi
\fi 

\ifnfsstwo
  \DeclareMathAlphabet{\mathbfit}{OT1}{cmr}{bx}{it}
  \SetMathAlphabet\mathbfit{bold}{OT1}{cmr}{bx}{it}
  \DeclareMathAlphabet{\mathbfss}{OT1}{cmss}{bx}{n}
  \SetMathAlphabet\mathbfss{bold}{OT1}{cmss}{bx}{n}
  \ifAMStwofonts
    \ifCUPmtlplainloaded \else
      \DeclareSymbolFont{UPM}{U}{eur}{m}{n}
      \SetSymbolFont{UPM}{bold}{U}{eur}{b}{n}
      \DeclareSymbolFont{AMSa}{U}{msa}{m}{n}
      \DeclareMathSymbol{\upi}{0}{UPM}{"19}
      \DeclareMathSymbol{\umu}{0}{UPM}{"16}
      \DeclareMathSymbol{\upartial}{0}{UPM}{"40}
      \DeclareMathSymbol{\leqslant}{3}{AMSa}{"36}
      \DeclareMathSymbol{\geqslant}{3}{AMSa}{"3E}
    \fi
  \fi
\fi 

\ifCUPmtlplainloaded \else
  \ifAMStwofonts \else 
    \def\upi{\pi}
    \def\umu{\mu}
    \def\upartial{\partial}
  \fi
\fi

\begin{document}
\title{High-resolution spectroscopy of V854\,Cen in decline -- Absorption
and emission lines of C$_2$ molecules}
\author[N. Kameswara Rao \& David L. Lambert]
       {N. Kameswara Rao,$^1$ David L. Lambert,$^2$\\ 
       $^1$Indian Institute of Astrophysics, Bangalore 560034, India\\
       $^2$Department of Astronomy, University of Texas, Austin, TX 78712-1083, USA\\}
\date{Accepted .
      Received ;
      in original form 1999 }

\pagerange{\pageref{firstpage}--\pageref{lastpage}}
\pubyear{1999}

\maketitle

\label{firstpage}

\begin{abstract}
High-resolution optical spectra of the R Coronae Borealis (RCB)
 star V854\,Centauri
in the early stages of a decline show, in addition to the features
reported for other  RCBs in decline, narrow absorption lines 
from the C$_2$ Phillips system. The low rotational temperature,
T$_{\rm rot}$ = 1150K, of the C$_2$ ground electronic state suggests
the cold gas is associated with the developing shroud of carbon dust.
These absorption lines were not seen at a fainter magnitude on the rise
from minimum light nor at maximum light. This is the first detection of
cold gas around a RCB star.

\end{abstract}


\section{Introduction}

R Coronae Borealis stars are H-poor F-G type supergiants that decline
in brightness unpredictably  by up to  8 magnitudes and
remain  below their normal brightness for several weeks to
months. It is generally accepted that these declines 
are due to formation of a cloud of carbon soot that obscures the
stellar photosphere. Unanswered questions  remain: `What triggers
cloud formation?' and `Where does the soot form?' High-resolution
spectroscopic monitoring of RCBs from maximum light into
decline will likely be necessary  to refine schematic
ideas into answers  that are
accorded widespread acceptance. We report 
the first detection of cool gas (T $\simeq 1100$K) during the early decline of 
a RCB star and, hence, evidence for a site of soot formation.
Cold dust is, of course, known around  RCBs 
through  detection of an infrared excess.

 The RCB in
question is V854\,Cen, which at maximum light is the
third brightest RCB variable after R\,CrB and RY\,Sgr.
V854\,Cen is presently  the most variable of all Galactic  RCBs.
Despite the combination of favorable apparent magnitude and propensity to
fade, there is a dearth of high-quality spectroscopic observations of
this star in decline. The sole report of a high-resolution optical
spectrum  covering a broad bandpass
in a deep decline is that by Rao \& Lambert (1993) taken
when the star had faded by about 8 mag. Low resolution
spectra are described by Kilkenny \& Marang (1989) and spectropolarimetric
observations are discussed by Whitney et al. (1992). 
Spectra at high-resolution at maximum light were used by Asplund et al (1998)
for their abundance analysis that confirmed that V854\,Cen has a somewhat
unusual composition among the RCBs for which abundance anomalies are
a {\it sine qua non}. In particular, V854\,Cen, although hydrogen-poor
relative to normal stars, is the most hydrogen-rich RCB by a clear margin.
Despite limited temporal coverage, our new  spectra of V854\,Cen in decline
provide a  novel result  - the detection of cold C$_2$ gas. 
Our  spectra  otherwise closely resemble
those of the RCBs extensively studied in decline:
R\,CrB (Rao et al. 1999) and RY\,Sgr (Alexander et al. 1972).
This concordance, which  suggests  that RCBs  have a common general
structure of their upper atmospheres and circumstellar
regions, is  briefly demonstrated here but we focus 
on the novel  lines of the C$_2$ molecule.

\section{Observations}

V854\,Cen was observed on four occasions 
from the W.J. McDonald Observatory with the
2.7m Harlan J. Smith reflector and the {\it 2dcoud\'{e}} spectrograph
(Tull et al. 1995). Details of the observations
are given in Table 1.  Figure 1 shows the light curve and the epochs of
our spectra. The first two spectra at
effectively the same epoch  were taken when the
star was at V $\sim$ 10.3  about 55 days after the onset of a decline
that  saw the star fade to V $\sim$ 13.6 by  1998 late-May.
We reobserved the star on 1998 June 6 at V $\sim$ 11.7 in its 
recovery to maximum brightness, and again on
1999 February 10 when the  star had returned to  maximum
brightness. Observations by amateur observers show that
the recovery from the deep decline in mid-1998 to maximum brightness in
early 1999 was rapid, unbroken by subsidiary declines, and faster
than the fall from maximum to minimum brightness which may have
been interrupted by brief halts.

\begin{table}
\centering
\begin{minipage} {140mm}
\caption{Spectroscopic observations of V854\,Centauri.}
\begin{tabular}{lrrrr} \hline
Date & {XJD}\footnote{XJD = JD - 245000.0}& V & {S/N}
\footnote{S/N ratio in continuum near 6560\AA.} &{Phase}%
 \footnote{Pulsation phase from Lawson et al.'s (1999) ephemeris\\ where zero phase is light maximum.} \\ \hline
1998 April 8   &    911.887&     10.3&      69&      81.227 \\
1998 April 10  &     913.832&    10.3&      53&      81.272\\
1998 June 6    &     970.679&    11.7&      61&      82.587\\
1999 Feb 10    &    1219.988&     7.3&     149&      88.354\\ \hline
\end{tabular}
\end{minipage}
\end{table}

\begin{figure}
\epsfxsize=8truecm
\epsffile{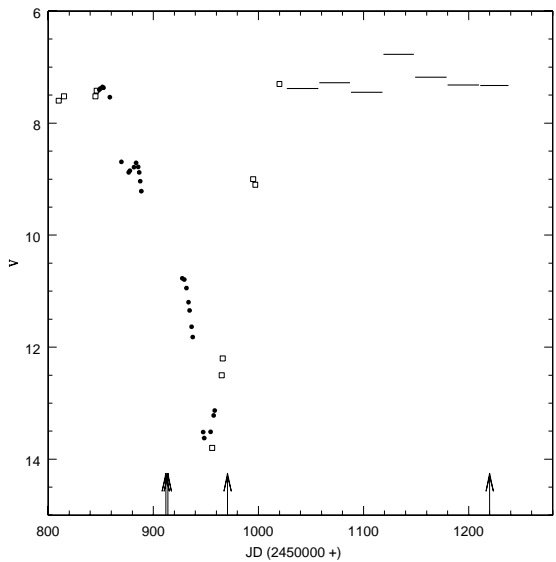}
\caption{The lightcurve for V854\,Cen from early 1998 to early 1999. Dots refer to
V magnitudes
 from Lawson et al. (1999), open squares and dashes are visual estimates from
the AAVSO. The Julian dates of our observations are
indicated by arrows.}
\end{figure}

The cross-dispersed echelle spectra are at a resolving power of 60,000 with nominal
coverage from 3800\AA\ to 10200\AA. Echelle orders are incompletely
captured on the CCD for wavelengths longward of 5500\AA. In addition,
the star's southerly declination (Dec. = - 39$^{\circ}$) and the observatory's
 northerly 
latitude  (Lat. = 31$^{\circ}$) result in
severe atmospheric dispersion and loss of signal in the blue such that
the spectra are not useable for wavelengths shorter than about 4100\AA.

\section{V854\,Centauri in decline}

Well-observed RCBs in decline -- R\,CrB and RY\,Sgr -- show common spectral
characteristics  that are shared with V854\,Cen. As a star fades,
the first prominent addition to its optical spectrum are two sets of
sharp emission lines:  E1 (Alexander et al. 1972) or `transient'
(Rao et al. 1999) appear shortly after onset of a decline and
disappear after a couple of weeks, and E2 or `permanent' lines
are prominent for a longer period and may be present in some or all
declines at all times, even at maximum light (Lambert, Rao, \& Giridhar 1990a).
 A 
mark of E1 lines is that they include high-excitation lines (C\,{\sc i},
O\,{\sc i}, and Si\,{\sc ii}, for example) not found among E2 lines.
Singly-ionised metals (e.g., Ti\,{\sc ii} and Fe\,{\sc ii}) are prominent
contributors of E1 and E2 lines.
The E1 and E2 lines are sharp (FWHM $\sim 14$ km s$^{-1}$ in R\,CrB). In
deep declines, a few broad emission lines are seen with FWHM $\sim 300$
km s$^{-1}$ with the Na\,D being the strongest. 

In our spectra of V\,854 Cen,
E1 lines, especially C\,{\sc i}
lines, are present in  1998
April: 46 C\,{\sc i} lines from 6400\AA\ to 8800\AA\ give a velocity of
-16.7 $\pm$ 2.8 km s$^{-1}$. Emission had gone by 1998 June with the same
lines appearing
 in absorption at  a velocity of -24.0 $\pm$ 2.4 km s$^{-1}$ which
is the (out-of-decline) mean velocity of -25 km s$^{-1}$ 
that  is maintained to
about 2 km s$^{-1}$ as the star undergoes small semi-regular
brightness variations (Lawson \& Cottrell 1989).
  The  velocity of infall of the C\,{\sc i}
emission lines is similar to that  seen for
R\,CrB (Rao et al. 1999). Lines of higher excitation such as the
N\,{\sc i} lines beyond 8000\AA\ also appear affected by emission, i.e.,
the N\,{\sc I} 8216\AA\ line has an equivalent width of 59 m\AA\ and a 
FWHM of 0.36\AA\ in April but its normal values, as in the 1999 February
spectrum, are 164 m\AA\ and 0.69\AA, respectively. It seems probable that
 emission has reduced the equivalent width, narrowed the line, and shifted
the apparent absorption velocity to the blue with the mean absorption
velocity at -34 $\pm$ 2 km s$^{-1}$ in 1998 April. Some C\,{\sc i}
emission lines show P Cygni profiles with absorption also at -34 km s$^{-1}$.
Emission from E1 and E2 lines affects almost all photospheric lines
in 1998 April. With the decay of E1 lines, the photospheric
velocity is measureable from  the 1998 June spectrum: the result
-27 $\pm$ 1 km s$^{-1}$ is consistent with the systemic velocity.
Lines of low and high excitation potential are at the systemic velocity on the
1999 February spectrum.

The E2 lines on the 1998 April and June spectra
are slightly blue-shifted with respect to the mean
photospheric velocity. The peak velocity which is unchanged between 
April and June  is -30 $\pm 1$ km s$^{-1}$ corresponding to a
blue shift of about 5 km s$^{-1}$, a typical value for the E2 lines
of R\,CrB and RY\,Sgr. The degree of excitation appears to be
similar to that of R\,CrB in its 1995-1996 decline, and the
line widths are also similar.
The line fluxes dropped by about a factor of 30 from 1998 April
to June, as the V flux dropped by a factor of only 4.  This contrasts
with the 1995-1996 decline of R\,CrB when the line fluxes dropped by
less than the V magnitude.

The only detectable broad lines are the Na\,D lines. Other
broad lines reported by Rao \& Lambert (1993) are not present. We 
attribute their absence to the fact that our observations were taken at
V = 11.7 (and 10.3) but the spectrum on which our earlier report
was based was obtained when the star was about 3 magnitudes fainter. 
Similarly, R\,CrB's broad lines  appeared only in the deepest part of
its decline.

Low-excitation lines of neutral metals are in absorption without discernible
emission but with their weak absorption red-shifted relative to the
systemic velocity: the mean velocity of +15 $\pm$ 2 km s$^{-1}$ from 7
lines on the 1998 April spectra implies infall at 40 km s$^{-1}$ relative
to the photosphere. Similar red-shifted
lines were seen in R\,CrB. This red-shifted
absorption, which is also clearly seen in the red wing of prominent
sharp (blue-shifted)  emission lines, is unlikely to be the residual
of the photospheric line (assumed to be at the systemic velocity) because
the red-shifted absorption occurs outside the normal photospheric profile and
many lines lack accompanying emission. The fact that the red-shifted
absorption appears in lines of different excitation potentials indicates 
that the responsible gas is warm.  By 1998 June, the same lines were
at -13 $\pm$ 2 km s$^{-1}$ and at the systemic velocity (i.e., photospheric
in origin)  by 1999 February.  

These snapshots of V\,854 Cen's spectrum  suggest its
decline from  onset to beyond minimum light largely behaved similarly
to R\,CrB's 1995-1996 decline.
There is one exciting novel feature revealed for V854\,Cen.

\section{C$_2$ Swan and Phillips System Lines}

 Previous detections of C$_2$ in spectra of
RCBs   are
for the Swan system which provides photospheric
 absorption lines  at maximum light in all but the hottest
RCBs, and  sharp and broad emission lines in decline spectra (Rao \&
Lambert 1993; Rao et al. 1999). Swan photospheric and E2 lines are seen here. 
The novel feature is the detection of low excitation  (non-photospheric)
 Phillips lines in absorption.

The Phillips system's lower state is the
 C$_2$ molecule's ground state (X\,$^1\Sigma^+_g$, Ballik \& Ramsay 1963;
Huber \& Herzberg 1979) and its   upper 
state (A\,$^1\Pi_u$) has the excitation energy T$_{\rm e}$ = 8391 cm$^{-1}$.
The Swan system's lower level is the lowest and very
low-lying  {\it triplet} state 
(a\,$^3\Pi_u$) with T$_{\rm e}$ = 716 cm$^{-1}$ and the upper state 
(d\,$^3\Pi_g$) is at T$_{\rm e}$ = 20022 cm$^{-1}$. Other low-lying 
states exist  but no other
band systems from the ground or low-lying states provide lines in our
bandpass. Radiative transitions between singlet and triplet states
occur with a low transition probability relative to the Phillips
singlet-singlet and Swan triplet-triplet transitions. 

Resolved rotational structure in the Swan system 0-0 band is shown in  Figure 2.
 The velocity,
as measured from clean 0-0 lines is -27 km s$^{-1}$ which is that of the
E2 sharp emission atomic lines. The line width, which is slightly greater than
the instrumental resolution, is
also equal to that of E2 atomic lines. The rotational temperature 
estimated following Lambert et al. (1990b) is T$_{\rm rot} = 4625 \pm $ 300K 
(see Figure 3).
Many bands from the $\Delta v$ = 0, $\pm$1, and +2 sequences are present.
 Semi-quantitative comparisons of the band profiles
in the $\Delta v$ = +1 sequence with predicted profiles (Lambert \& Danks
1983) indicate a vibrational temperature near 5000K and, hence,
likely equal to the rotational temperature. Rao \& Lambert (1993 - see
also Rao et al. 1999 for R\,CrB) in a deeper decline found the Swan lines
to be broad but in our spectra any broad component must be very weak.

\begin{figure}
\epsfxsize=8truecm
\epsffile{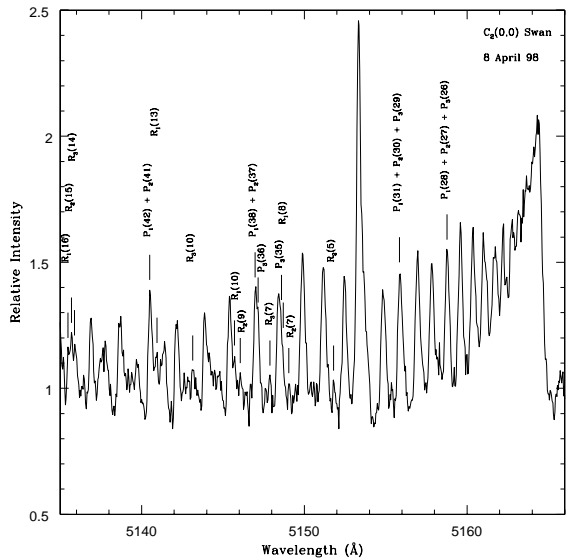}
\caption{The C$_2$ Swan 0-0 P branch bandhead on 1998 April 8. A few lines and
blends of are identified.}
\end{figure}

\begin{figure}
\epsfxsize=8truecm
\epsffile{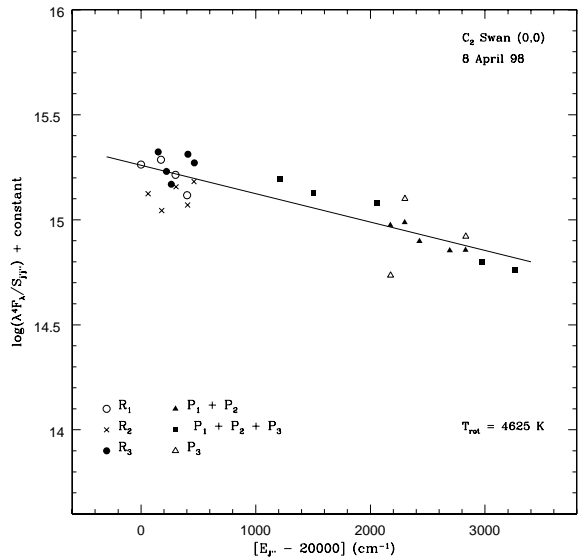}
\caption{A standard Boltzmann plot compiled from Swan 0-0 lines and blends
(see key on figure). The solid line is a least-squares fitted line corresponding
to a rotational temperature of 4625K.}
\end{figure}

Weak absorption lines identified as Phillips system lines are present
on the 1998 April spectra but absent from the 1998 June spectrum. Figure 4 
shows a portion of the 2-0 band and includes a spectrum of the post-AGB 
star IRAS 22223+4327 in which circumstellar C$_2$ lines
are strong. Many lines from the 2-0 and 3-0 bands were detected in
V854\,Cen with equivalent widths of up to 50 m\AA.
 A search for 3-1 and 4-1 lines was unsuccessful; this is  not
surprising given the low excitation temperature found from the detected
lines. No search was made for either 1-0 or 4-0 lines. Rest wavelengths
from Bakker et al. (1997) give the radial velocity of 
-30.4 $\pm$ 1.3 km s$^{-1}$ from 15 lines,
 i.e., a  small expansion velocity relative to the
systemic velocity of -25 km s$^{-1}$. The velocity
differs considerably from that (+15 km s$^{-1}$)
 of the red-shifted absorption component
of low-excitation atomic lines.
 In contrast to photospheric lines,
the C$_2$ absorption lines are not resolved. 
Boltzmann plots for 2-0 and 3-0 lines give a mean rotational temperature 
of T$_{\rm rot} = 1150 \pm$ 70K from levels J$^{\prime\prime}$ =
4 to 28. We interpret this as a close approximation to the
gas kinetic temperature. If, as occurs in interstellar diffuse clouds,
the C$_2$ molecule's excitation is greatly influenced by
radiative pumping in the Phillips bands (and X\,$^1\Sigma^+_g$
 $\rightleftharpoons$ a\,$^3\Pi_u$ radiative transitions), the
Boltzmann plot is expected to be curved with lowest rotational levels
giving a temperature close to the kinetic temperature and higher levels
a higher temperature dependent on the ratio of the gas density and
the photon flux in the near-infrared, for example, the ground state
populations for $\zeta$\,Oph's diffuse clouds give T$_{\rm rot} \simeq
40$K from the lowest levels and 785K from levels J$^{\prime\prime}
\simeq 20$ (Lambert, Sheffer \& Federman 1995). A linear Boltzmann
plot, as here, suggests that the observed levels may be in
equilibrium with the gas, i.e., our C$_2$ molecules are in 
gas at a temperature below that  at which carbon dust grains  form,
and the molecules may well be mixed in with the fresh dust. 
The molecular column density is about 2 $\times 10^{15}$ cm$^{-2}$.

\begin{figure}
\epsfxsize=8truecm
\epsffile{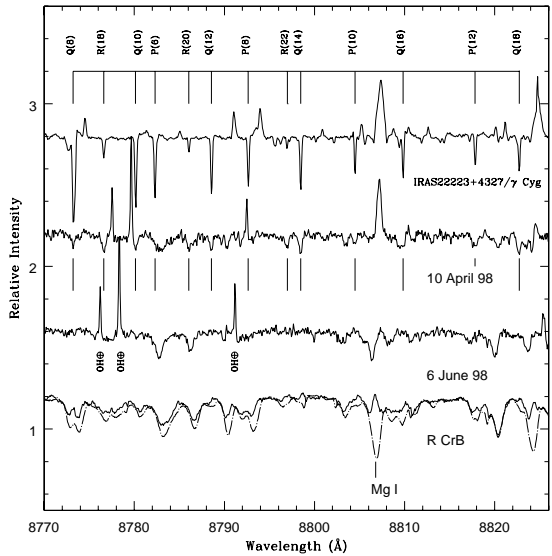}
\caption{Spectra from 8770-8826\AA\ of V854\,Cen and IRAS2223+4327.
Locations of C$_2$ Phillips 2-0 lines are indicated at the top of the
figure and below the 1998 April 10 spectrum of V854\,Cen. The Phillips
lines are strongly present in IRAS22223+4327 and the 1998 April spectra
of V854\,Cen but not in the 1998 June 6 spectrum. Two R\,CrB spectra
are shown superimposed: the spectrum from 1995 September 30 (dash-dot
line) was taken at
maximum light just prior to the 1995 decline;
 the spectrum from 1995 October 13 (solid line) was taken when the star was about 3 magnitudes below maximum bightness, i.e., the star had faded by about the
same amount as V854\,Cen had on 1998 April 10.}
\end{figure}

\begin{figure}
\epsfxsize=8truecm
\epsffile{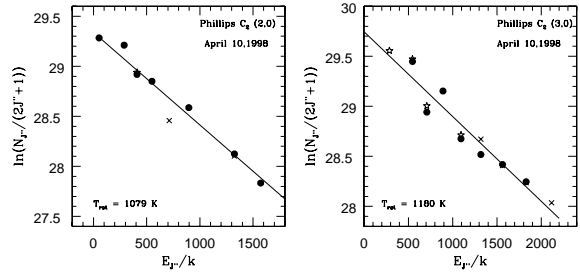}
\caption{Standard Boltzmann plots for the C$_2$ Phillips absorption
lines. Stars, dots and crosses refer to P, Q and R branch lines, 
respectively. The line is the least-squares fitted
line corresponding to the rotational temperature indicated on the
figure.}
\end{figure}

Five questions arise directly from these observations of C$_2$ lines:
Why is an absorption component not seen in the  C$_2$ Swan lines? Why is
an emission component not seen in the C$_2$ Phillips lines? How are the
Swan emission lines excited? Where is the emitting gas? Where is the
cold absorbing gas?

The apparent absence of Swan absorption lines is easily explained. In the
weak-line limit, the equivalent width 
$W_{\lambda} \propto f\lambda^2NL$ where $f$ and $\lambda$ are
 the line's oscillator strength and wavelength respectively and $NL$ is the
column density of molecules in the lower level of the transition. If the
column densities in the lowest singlet and triplet states are equal, the
Swan system lines are favored by a factor of about 7 with the system's
greater $f$-value being a major factor (Grevesse et al. 1991; Bakker \&
Lambert 1998) but considering that the Phillips absorption lines
are at almost the same velocity as the Swan emission lines (-30 km s$^{-1}$
versus -27 km s$^{-1}$), we suppose that the Swan absorption
lines are masked by the strong emission lines. A large
increase in the column density of the lowest triplet state relative to
the ground triplet state would be required to provide detectable
absorption.  

A plausible explanation may be offered for the absence of Phillips
emission lines. An approximate flux calibration of our spectrum gives
the detection limit for a sharp Phillips system line at about 0.2 that
of a single sharp Swan line.
The predicted relative fluxes in Swan and Phillips
lines depends on the assumed mode of excitation. If the molecule is 
in thermal equilibrium at the measured T$_{\rm rot}$ = 4625K,
it is readily shown that the flux in a Phillips 2-0  line  is about 15\%
that of the Swan 0-0 of a similar J-value in the event that
reddening may be neglected, i.e., the line would not
appear in emission in our spectrum. The great difference in the $f$-values of
the 0-0 Swan band and the 2-0 Phillips band is a major contributor to the
low flux of the Phillips lines. In the case of resonance
fluorescence, as occurs for comets, the Phillips line is similarly
weak unless the population in the X\,$^1\Sigma^+_g$ state is very
much greater than in a\,$^3\Pi_u$ state. At low particle densities,
as in the interstellar medium, the latter state is not populated;
electric-quadrupole transitions drain population to the lowest 
levels of the X state. This situation is, however, unlikely to
prevail in V854\,Cen. For R\,CrB, which may be taken as similar to
V854\,Cen, the sharp emission lines come from a region of high
particle density (Rao et al. 1999) such that the a to X populations
must be close to their equilibrium value, i.e., sharp Phillips emission lines
are almost certainly below our detection limit. 

Rao et al. (1999) assembled a wealth of data on the E2 lines including
C$_2$ Swan system lines  seen
throughout R\,CrB's 1995-1996 decline to determine that the emitting gas
was warm and dense. The location of the gas relative to the star and the
obscuring dust cloud could not be definitively established.
 Similarities between the atomic E2 lines of the two stars strongly suggest that
 V854\,Cen's
E2 Swan lines originate in the region providing its atomic E2 lines, and
that this region resembles that around R\,CrB. Differences in physical
condition and chemical composition may account for the fact that
the Swan lines are more strongly in emission from V854\,Cen.  
 
Our temporal coverage of V854\,Cen's decline is limited but encourages
the speculations that (i) the narrow C$_2$ absorption lines appear
only in decline, and  (ii) the appearance of the E1 (transient) atomic
 lines and the
C$_2$ absorption lines may be related. The C$2$ Swan lines appear as
weak photospheric absorption lines at maximum light. Their Phillips
system counterparts are too weak to detect. The 1100K absorption
lines are absent from our 1999 February spectrum. 
 The E1 lines and
Phillips absorption are both present in 1998 April but  are
seen neither in 1998 June when the star was fainter and
 E2 lines remained prominent.
 This suggests that the 1100K absorption
is not merely related to the dust but more to the early
stages of the decline. A possible connection is the
presence of a shock, as considered  by Woitke, Goeres \& Sedlmayr
(1996) and Woitke (1997) to be the trigger for a RCB decline. In their
scenario,  E1 lines originate in the hot gas immediately
behind the shock and dust forms in cool dense gas further behind the outwardly
moving shock front. Here, as for R\,CrB, the shock may propagate through
the infalling gas betrayed by the red-shifted absorption lines of low
excitation atomic lines. 

 In light of the detection of the Phillips absorption lines,
 we have reexamined spectra
 of R\,CrB obtained in its 1995-1996 decline (Rao et al.
1999). R\,CrB appears not to have shown these absorption lines in
its decline (Figure 4) but did show the E1 high-excitation lines.
This difference between V854\,Cen and R\,CrB may reflect
differences in physical conditions
in the upper atmospheres or in chemical compositions. Such differences
may also account for the far greater propensity of V854\,Cen to
go into decline.
The high hydrogen to carbon ratio of V854\,Cen 
has led Goeres (1996) to predict that formation of carbon-containing molecules
and dust grains is controlled by acetylene (C$_2$H$_2$) rather than the
 C$_3$ molecules that act as the throttle for `normal' RCBs.

\section{Concluding Remarks}

For the first time, cold gas below the temperature required for
soot formation has been detected around a RCB in its decline to
a deep minimum. Our detection of absorption by cold C$_2$ molecules
around V854\,Cen now needs to be followed by synoptic observations
of this active RCB variable in order to trace the evolution of the
 cold gas and to place it relative to the star. Is the gas located behind
a shock where, as some theories would suppose, dust formation is
triggered? Or is it merely an innocent companion to the dust? We
recognise that providing synoptic observations at an adequate
spectral resolution and high temporal frequency is  a 
substantial challenge. If the challenge can be met, the result
will be a window into the time and place of dust formation, and,
perhaps, provie the long-sought understanding of how
 the characteristic declines of RCBs
are initiated.

\section{Acknowledgements}
We would like to thank Jocelyn Tomkin and Gulliermo Gonzalez for obtaining
the  spectra of V854\,Cen in decline, and Gajendra Pandey for
considerable help in the reduction and presentation of the
spectra.
This research has been supported in part by the US National Science
Foundation (grant AST 9618414.

\label{lastpage}

\end{document}